\documentclass[a4paper,12pt]{article}
\usepackage[a4paper,text={16.8cm,22.4cm}]{geometry}
\usepackage{amsmath,amssymb,bm,psfrag,graphicx,color}
\allowdisplaybreaks 
\begin{document}

\begin{titlepage}

\begin{flushright}
MZ-TH/11-08\\
April 20, 2011
\end{flushright}

\vspace{0.2cm}
\begin{center}
\Large\bf
{\boldmath
Factorization and Resummation for Jet Broadening
\unboldmath}
\end{center}

\vspace{0.2cm}
\begin{center}
Thomas Becher$^a$, Guido Bell$^a$ and Matthias Neubert$^b$\\
\vspace{0.4cm}
{\sl 
${}^a$\,Institut f\"ur Theoretische Physik, Universit\"at Bern\\
Sidlerstrasse 5, CH--3012 Bern, Switzerland\\[0.3cm]
${}^b$\,Institut f\"ur Physik (THEP), 
Johannes Gutenberg-Universit\"at\\ 
D--55099 Mainz, Germany}
\end{center}

\vspace{0.2cm}
\begin{abstract}
\vspace{0.2cm}
\noindent 
Jet broadening is an event-shape variable probing the transverse momenta of particles inside jets. It has been measured precisely in $e^+ e^-$ annihilations and is used to extract the strong coupling constant. The factorization of the associated cross section at small values of the broadening is afflicted by a collinear anomaly. Based on an analysis of this anomaly, we present the first all-order expressions for jet-broadening distributions, which are free of large perturbative logarithms in the two-jet limit. Our formulae reproduce known results at next-to-leading logarithmic order but also extend to higher orders.
\end{abstract}
\vfil

\end{titlepage}

\section{Jet broadening}

Event shapes measure geometric properties of collider events. They are conceptually simpler than the more commonly used jet observables and have been measured with exquisite precision at lepton colliders such as LEP \cite{Heister:2003aj,Abdallah:2004xe,Achard:2004sv,Abbiendi:2004qz}. By comparing to theoretical predictions, these results can be used to extract the strong coupling constant and to search for new particles. The interest in event shapes has been renewed recently after the fixed-order results at next-to-next-to-leading order (NNLO) in QCD perturbation theory have become available \cite{GehrmannDeRidder:2007hr,Weinzierl:2009ms}. However, even with these corrections at hand, the perturbative uncertainty still dominates the error on the extracted value of $\alpha_s$. The convergence of the perturbative series can be improved by resumming those perturbative corrections which become dominant near kinematical end-points. This resummation was performed at next-to-leading logarithmic (NLL) accuracy for the event shapes measured at LEP; however, to improve the accuracy of $\alpha_s$ it is necessary to also resum subleading logarithms of higher order. This has been achieved for the event-shape variables thrust \cite{Becher:2008cf} and heavy-jet mass \cite{Chien:2010kc}. In the present paper we derive, for the first time, all-order formulae for the total and wide jet broadenings, which are free of large logarithms and allow the resummation of higher logarithms also for these quantities. 

The jet broadening is defined as follows. With each hadronic event in an $e^+ e^-$ collision one associates a thrust axis $\vec{n}_T$, which is the direction of maximum three-momentum flow. The particles of the event can then be divided into two groups: those moving in the forward hemisphere with respect to the thrust direction ($\vec{p}_i\cdot\vec{n}_T>0$), and those moving in the opposite hemisphere. For simplicity, we assume that $\vec{n}_T$ points to the left and refer to the two groups of particles as the left-moving and right-moving ones, respectively. The left-broadening is defined as the sum of the absolute values of the transverse momenta of the left-moving particles \cite{Rakow:1981qn,Ellis:1986ig}
\begin{equation}
   b_L = \frac{1}{2} \sum_{i\in L} |{\vec{p}_i}^\perp| 
   = \frac{1}{2} \sum_{i\in L} |\vec{p}_i \times \vec{n}_T| \,, 
\end{equation}
and analogously one defines the right-broadening $b_R$. Usually, one normalizes the broadenings to the center-of-mass energy $Q=\sqrt{s}$ and defines $B_{L,R}=b_{L,R}/Q$, but we prefer to work with the dimensionful quantities $b_{L,R}$. What is measured experimentally is the total broadening $b_T=b_L+b_R$ as well as the wide broadening $b_W=\max(b_L,b_R)$. 

In the following, we are interested in the region of small but perturbative broadenings, $\Lambda_{\rm QCD} \ll b_L \sim b_R \ll Q$, where $\Lambda_{\rm QCD}$ is a typical scale associated with non-perturbative strong-interaction physics. Despite the fact that the broadenings are in the perturbative domain, fixed-order perturbation theory breaks down at small values of $b_{L,R}$, because large logarithms arise. For the fraction of events with total broadening smaller than $b_T$, for example, one obtains at next-to-leading order, 
\begin{equation}\label{RB}
   R(B_T) = \frac{1}{\sigma_0} \int_0^{B_T Q}\!db_T\,\frac{d\sigma}{db_T} 
   = 1 + \frac{C_F\alpha_s}{2\pi} \left( -4\ln^2 B_T - 6\ln B_T - 7 + \pi^2 \right),
\end{equation}
up to terms which vanish in the limit where $B_T=b_T/Q$ goes to zero. Here and below we normalize to the tree-level total cross section $\sigma_0$. In the result (\ref{RB}) the Sudakov logarithms are manifest. To obtain reliable predictions for small broadening, these enhanced corrections must be resummed to all orders. At leading double-logarithmic order this was achieved in \cite{Catani:1992jc}. To this accuracy the broadening can be written as a product of two jet functions in Laplace space. An improved version of this result, valid also at the single logarithmic level, was later presented in \cite{Dokshitzer:1998kz}. However, an all-order formula for broadening which is free of large logarithms was missing. Near the two-jet limit, event shapes such as thrust factorize into a convolution of a hard function, two jet functions, and a soft function, and this factorization forms the basis for an all-order resummation of logarithmically enhanced corrections. This was shown in \cite{Berger:2003iw} for a large class of event-shape variables, but it was also pointed out that this factorization breaks down for broadening. The same class of event shapes was recently reanalyzed in Soft-Collinear Effective Theory (SCET) in \cite{Hornig:2009vb}, which concluded that the usual effective-theory power counting breaks down for broadening.

While the naive soft-collinear factorization indeed breaks down for broadening, we nevertheless manage to derive in this paper all-order formulae for the total and wide broadening distributions, which are free of large logarithms. The reason is that the breaking of factorization has a very specific origin. In the effective theory, it manifests itself as a collinear anomaly, which generates an additional dependence on the large momentum transfer $Q$ in the product of the jet and soft functions. In the effective theory the collinear anomaly is a quantum anomaly in the usual sense, that a symmetry of the classical (effective) Lagrangian is not preserved by the regularization. The factorization analysis is similar to the one for small-$q_T$ resummation in Drell-Yan production \cite{Becher:2010tm}. As in this case, in an intermediate step one needs to introduce additional regulators beyond dimensional regularization in order to obtain well-defined expressions in the effective theory. When the regulators are removed in the final predictions for physical cross sections, the anomalous $Q$ dependence remains. The regulator independence of the product of jet and soft functions gives a strong constraint on the dependence of the individual functions on $Q$, implying that this dependence must exponentiate \cite{Becher:2010tm,Chiu:2007dg}.

The analysis of jet broadening is complicated by the fact that not only collinear modes, but also a soft mode (using SCET$_{\rm II}$ terminology), whose momentum components scale as $p_s^\mu\sim b_T$, give a leading contribution to the cross section. In contrast to the present case, this soft mode does not contribute to $q_T$ resummation in Drell-Yan production at small transverse momentum, because the corresponding loop integrals are scaleless and can be omitted after proper regularization \cite{Becher:2010tm}. This is no longer the case for jet broadening, since the radiation is restricted to one of the hemispheres. Interestingly, we verified that starting at two-loop order the ultra-soft momentum region $p_{us}^\mu\sim b_T^2/Q$ also gives non-vanishing contributions to individual diagrams in the presence of the hemisphere constraint. However, as explained in \cite{Becher:2010tm}, in the sum of all graphs these contributions cancel as a result of the KLN theorem, and consequently the ultra-soft region does not contribute to the broadening.

\begin{figure}
\begin{center}
\psfrag{L}{$L$}\psfrag{R}{$R$}\psfrag{n}{$\vec{n}_T$}
\includegraphics[width=0.62\textwidth]{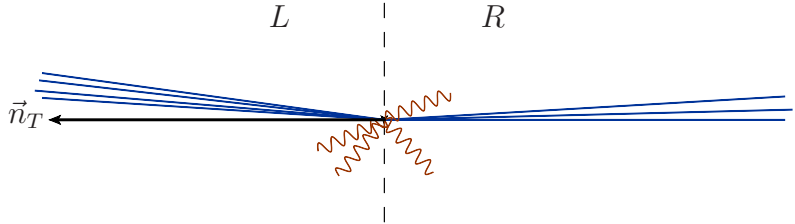}
\end{center}
\vspace{-0.5cm}
\caption{\label{fig:broadening}
A typical event with small broadening consists of energetic collinear partons in each hemisphere (blue lines) accompanied by soft radiation (red wiggly lines). The total transverse momentum with respect to the thrust axis $\vec{n}_T$ vanishes in each hemisphere.}
\end{figure}

We have stressed that the jet and soft functions relevant for broadening are not well defined without additional regularization. Leaving this issue aside for the moment, the naive factorization theorem for small broadening has the form
\begin{equation}
\begin{aligned}\label{naivefac}
   \frac{1}{\sigma_0}\,\frac{d^2\sigma}{db_L\,db_R} 
   &= H(Q^2,\mu) \int\!db_L^s \int\!db_R^s \int\!d^{d-2}p^\perp_L \int\!d^{d-2}p^\perp_R \\
   &\quad\times {\cal J}_L(b_L-b_L^s,p_L^\perp, \mu)\, {\cal J}_R(b_R-b_R^s,p_R^\perp,\mu)\, 
    {\cal S}(b_L^s,b_R^s,-p_L^\perp, -p_R^\perp,\mu) \,.
\end{aligned}
\end{equation}
The convolutions over $b_{L,R}^s$ arise because the physical broadening is the sum of the collinear and soft broadenings. The definition of the thrust axis ensures that the total transverse momentum  vanishes in each hemisphere, so if the left-moving collinear partons have transverse momentum $p_L^\perp$, the transverse momentum of the soft partons in the left hemisphere must be equal and opposite (see Figure~\ref{fig:broadening}).

The hard function $H(Q^2,\mu) =|C_V(-Q^2-i\varepsilon,\mu)|^2$ is the square of the quark vector form factor, and is known to three-loop accuracy \cite{Baikov:2009bg,Gehrmann:2010tu}. The quark jet function for the left-moving collinear partons is given by
\begin{equation}
\begin{aligned}
   \frac{\pi}{2}\,({n\!\!\!/})_{\alpha\beta}\,{\cal J}_L(b,p^\perp,\mu) 
   &= \sum\hspace{-0.55cm}\int\limits_X\,\,(2\pi)^d\,\delta(\bar{n}\cdot p_X-Q)\, 
    \delta^{d-2}(p_X^\perp-p^\perp) \\[-2mm]
   &\quad\times \delta\Big(b-\frac{1}{2} \sum_{i\in X} |p_i^\perp|\Big)\,   
    \langle 0|\chi_\alpha(0)|X\rangle\,\langle X|\bar{\chi}_\beta(0)|0\rangle \,,
\end{aligned}
\end{equation}
where $n^\mu=(1,\vec{n}_T)$ is a light-like vector along the thrust axis, $\bar{n}^\mu=(1,-\vec{n}_T)$ is its conjugate, and for simplicity we drop the subscript $L$ on the variables of the jet function. The first two $\delta$-distributions ensure that the produced jet $X$ has the desired energy and that its total transverse momentum has a given value $p^\perp$. In the absence of soft-collinear interactions, the collinear SCET Lagrangian is equivalent to QCD and the collinear quark field $\chi(x)$ can be identified with $\chi(x)=\frac{n\!\!\!/\bar n\!\!\!/}{4}\,W^\dagger(x)\,\psi(x)$, where $\psi(x)$ is the QCD quark field and $W(x)$ a straight Wilson line along the $\bar{n}^\mu$ direction from $-\infty$ to $x$ (see e.g.\ \cite{Becher:2009qa} for more details). In our computation of the jet function, we will use the standard QCD Lagrangian and Feynman rules. The jet function ${\cal J}_R$ for the right-moving collinear partons is obtained by exchanging $n^\mu\leftrightarrow\bar{n}^\mu$ in the above formula. The soft function is obtained as
\begin{equation}\label{softfun}
\begin{aligned}
   {\cal S}(b_L,b_R,p_L^\perp,p_R^\perp,\mu) 
   &= \sum\hspace{-0.825cm}\int\limits_{X_L,X_R}\!\delta^{d-2}(p_{X_L}^\perp-p_L^\perp)\, 
    \delta^{d-2}(p_{X_R}^\perp-p_R^\perp) \\
   &\quad\times \delta\Big(b_L-\frac{1}{2} \sum_{i\in X_L} |p_{L,i}^\perp|\Big)\, 
    \delta\Big(b_R-\frac{1}{2} \sum_{j\in X_R} |p_{R,j}^\perp|\Big) 
    \left| \langle X_L\,X_R| S_n^\dagger(0)\,S_{\bar{n}}(0) |0\rangle \right|^2 .
\end{aligned}
\end{equation}
Here $S_n$ and $S_{\bar n}$ are soft Wilson lines extending along the $n^\mu$ and $\bar n^\mu$ directions. The final states are split into left and right-moving particles, $X=X_L+X_R$, which contribute to the respective broadenings.

Instead of working with the cross section in momentum space, it is more convenient to discuss the Laplace-transformed cross section
\begin{equation}
   \frac{d^2\sigma}{d\tau_L\,d\tau_R} 
   = \int_0^\infty\!db_L\,e^{-\tau_L b_L} \int_0^\infty\!db_R\,e^{-\tau_R b_R}\,
    \frac{d^2\sigma}{db_L\,db_R} \,.
\end{equation}
It is furthermore beneficial to Fourier transform in the momenta $p_L^\perp$ and $p_R^\perp$, after which the factorization theorem (\ref{naivefac}) takes the form
\begin{equation}\label{ff1}
\begin{aligned}
   \frac{1}{\sigma_0}\,\frac{d^2\sigma}{d\tau_L\,d\tau_R} 
   &= (2\pi)^{2(d-2)}\,H(Q^2,\mu) \int\!d^{d-2}x_L^\perp \int\!d^{d-2}x_R^\perp \\
   &\quad\times \widetilde {\cal J}_L(\tau_L,x_L^\perp,\mu)\,
    \widetilde{\cal J}_R(\tau_R,x_R^\perp,\mu)\,
    \widetilde{\cal S}(\tau_L,\tau_R,x_L^\perp,x_R^\perp,\mu) \,,
\end{aligned}
\end{equation}
where $\widetilde {\cal J}_L$, $\widetilde {\cal J}_R$, and $\widetilde {\cal S}$ are the Laplace and Fourier-transformed jet and soft functions, e.g.\ 
\begin{equation}
   \widetilde{\cal J}_L(\tau_L,x_L^\perp,\mu)
   = \int_0^\infty\!db_L\,e^{-\tau_L b_L} \int\frac{d^{d-2}p_L^\perp}{(2\pi)^{d-2}}\,
    e^{-ip_L^\perp\cdot x_L^\perp}\,{\cal J}_L(b_L,p_L^\perp,\mu) \,.
\end{equation}
Since by rotational invariance the jet functions only depend on the moduli $|x_{L,R}^\perp|$ of the transverse-position vectors, we can integrate the soft function over the solid angles associated with these vectors. Introducing the dimensionless variables $z_{L,R}=2|x_{L,R}^\perp|/\tau_{L,R}$, the naive factorization formula then takes the final form
\begin{equation}\label{factfinal}
   \frac{1}{\sigma_0}\,\frac{d^2\sigma}{d\tau_L\,d\tau_R} 
   = H(Q^2,\mu) \int_0^\infty\!dz_L \int_0^\infty\!dz_R\,
    \overline {\cal J}_L(\tau_L,z_L,\mu)\,\overline{\cal J}_R(\tau_R,z_R,\mu)\, 
    \overline{\cal S}(\tau_L,\tau_R,z_L, z_R,\mu) \,, 
\end{equation}
with
\begin{equation}
\begin{aligned}
   \overline{\cal S}(\tau_L,\tau_R,z_L, z_R,\mu) 
   &= \frac{1}{{\cal N}^2} \int\!d\Omega^L_{d-2} \int\!d\Omega^R_{d-2}\, 
    \widetilde{\cal S}(\tau_L,\tau_R,x_L^\perp,x_R^\perp,\mu) \,, \\
   \overline{\cal J}_{L,R}(\tau,z,\mu) 
   &= {\cal N}\,(2\pi)^{d-2}\,\frac{\tau}{2} \left(\frac{\tau z}{2}\right)^{d-3}\, 
    \widetilde{\cal J}_{L,R}(\tau,x^\perp,\mu) \,.
\end{aligned}
\end{equation}
The normalization factor
\begin{equation}
   {\cal N} = \frac{\Omega_{d-2}}{(2\pi)^{d-2}}
   = \frac{2}{(4\pi)^{1-\epsilon}\,\Gamma(1-\epsilon)}
\end{equation}
is chosen such that for a $d=4-2\epsilon$ dimensional space-time the soft function $\overline{\cal S}^{(0)}=1$ at lowest order. For the lowest-order jet functions, we obtain
\begin{equation}\label{treejet}
   \overline{\cal J}_{L,R}^{(0)}(\tau, z)  
   = \frac{4^{\epsilon}\,\Gamma(2-2\epsilon)}{\Gamma^2(1-\epsilon)}\,
    \frac{z^{1-2\epsilon}}{\left(1+z^2\right)^{3/2-\epsilon}} \,.
\end{equation}

\section{Evaluation of the jet  and soft functions}

As mentioned above, the Feynman diagrams contributing to the jet and soft functions are not well defined individually in dimensional regularization. To evaluate them, we introduce additional analytic regulators \cite{Smirnov:1993ta,Smirnov:2002pj} in the QCD diagrams contributing to the broadening. This is done as follows. Let us assume that the QCD vector current produces a quark which ends up in the left-moving jet, while the anti-quark ends up in the right-moving jet. We now raise all propagators along the quark line to a fractional power,
\begin{equation}\label{analytreg}
   \frac{1}{p^2+i\varepsilon} 
   \to \frac{(\nu_1^2)^{\alpha}}{\left(p^2+i\varepsilon\right)^{1+\alpha}} \,,
\end{equation}
and analogously for the anti-quark line, but with a regulator $\beta$ and an associated scale $\nu_2$. Since the QCD diagrams are well defined without additional regulators, this operation is trivial for the full theory, but with the analytic regulators in place the effective-theory diagrams are now also well defined. The individual jet and soft functions in the naive factorization formula (\ref{factfinal}) exhibit poles in the regulators $\alpha$ and $\beta$, which only cancel in the product of these functions entering the formula for the cross section. As is explained in detail in \cite{Becher:2010tm}, the regulator on the anti-quark line regularizes the Wilson line in the collinear jet function of left-moving particles. The usual eikonal Feynman rule for the emission of a left-moving collinear gluon with momentum $k$ from the Wilson line $W^\dagger$ in the SCET quark field $\chi$ is modified to
\begin{equation}\label{wilsonreg}
   \frac{\bar{n}^\mu}{\bar{n}\cdot k} 
   \to \frac{\left(\nu_2^2\right)^\beta \bar{n}^\mu\,n\cdot p_R}%
            {\left(\bar n\cdot k\,n\cdot p_R\right)^{1+\beta}} \,,
\end{equation}
where $p_R$ is the total {\em right-moving\/} momentum, which fulfils $n\cdot p_R=Q$ at leading power in $b_{T}/Q$. Similarly, the rule for the emission of a right-moving collinear gluon from the corresponding Wilson line is obtained by replacing $\nu_2\to\nu_1$, $\beta\to\alpha$, $n^\mu\leftrightarrow\bar n^\mu$, and $p_R\to p_L$ in the above result. Note that the usual eikonal identities no longer hold in the presence of the analytic regulators. The Feynman rules for multiple emissions from the regularized object $W^\dagger$ are thus more complicated than for standard Wilson lines, but they can easily be derived by considering the corresponding regularized QCD diagrams.

The regulators $\alpha$ and $\beta$ also regularize the soft emission diagrams in the effective theory. The eikonal factor relevant for gluon emission from the regularized soft Wilson line $S_n$ is replaced by
\begin{equation}\label{softwilsonreg}
   \frac{n^\mu}{n\cdot k}\to 
   \frac{\left(\nu_1^2\right)^\alpha n^\mu\,\bar n\cdot p_L}%
        {\left(n\cdot k\,\bar n\cdot p_L\right)^{1+\alpha}} \,.
\end{equation}
The same expression with $\nu_1\to\nu_2$, $\alpha\to\beta$, $n^\mu\leftrightarrow\bar n^\mu$, and $p_L\to p_R$ holds for $S_{\bar n}$, the Wilson line describing soft emissions from the anti-quark.

\begin{figure}
\begin{center}
\begin{tabular}{ccccccc}
\includegraphics[width=0.19\textwidth]{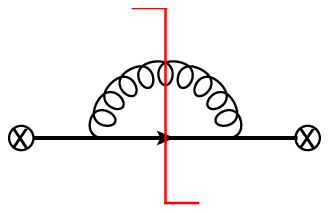} & & 
\includegraphics[width=0.19\textwidth]{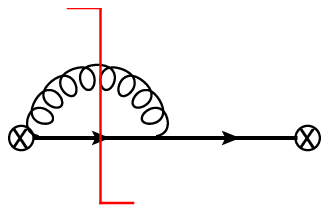} & &
\includegraphics[width=0.19\textwidth]{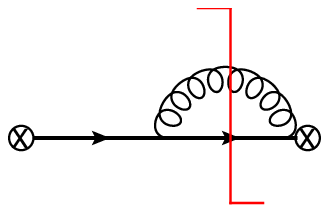} & & 
\includegraphics[width=0.19\textwidth]{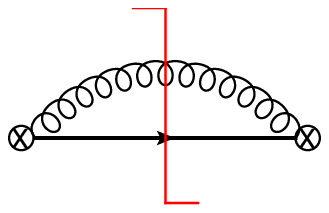}
\end{tabular}
\end{center}
\vspace{-0.5cm}
\caption{\label{fig:oneloop}
Next-to-leading corrections to the jet function. The one-loop virtual corrections are scaleless and vanish.}
\end{figure}

With the additional regulators in place, we may now evaluate the next-to-leading order  corrections to the jet and soft functions. The Feynman diagrams relevant for the jet function are shown in Figure~\ref{fig:oneloop}. Let us first consider the simple case $p_{L,R}^\perp=0$, for which the one-loop expression for the jet function can be obtained analytically. In terms of the $\overline{\rm MS}$ coupling constant $\alpha_s\equiv\alpha_s(\mu)$, the next-to-leading order result for the bare diagrams relevant for the left-moving jet reads
\begin{equation}\label{oneloopJ}
\begin{aligned}
   {\cal J}_L(b,p^\perp=0) 
   &= \delta(b) + \frac{C_F\alpha_s}{2\pi}\,\frac{e^{\epsilon\gamma_E}}{\Gamma(1-\epsilon)}\,
    \frac{1}{b} \left( \frac{\mu}{b} \right)^{2\epsilon} \\
   &\quad\times \left[ (1-\epsilon )
    + \frac{4\Gamma(2+\alpha)\,\Gamma(\alpha-\beta)}{\Gamma(2+2\alpha-\beta)}
    \left( \frac{\nu_1^2}{b^2} \right)^{\alpha} \left( \frac{\nu_2^2}{Q^2} \right)^{\beta} 
    \right] .
\end{aligned}
\end{equation}
The first diagram in Figure~\ref{fig:oneloop} produces the first term inside the brackets, which is  well defined without analytic regulators and thus has been evaluated setting $\alpha=\beta=0$. The second and third diagrams give identical contributions and produce the second term inside the brackets. These graphs are only well defined if the analytic regulators fulfill $\alpha\neq\beta$. One can thus only set one of the regulators to zero. Expanding then in the second regulator one encounters a pole. The fourth diagram vanishes. Note that at next-to-leading order the regularized jet function is gauge invariant, since only real-emission diagrams contribute. At higher orders, the two jet functions and the soft function will likely not be gauge invariant individually, since analytic regularization breaks gauge invariance; however, gauge invariance will be recovered in the product of the three functions. The result for the jet function ${\cal J}_R$ follows from the expression for ${\cal J}_L$ given above by exchanging $\alpha\leftrightarrow\beta$ and $\nu_1\leftrightarrow\nu_2$. For $p_{L,R}^\perp\neq 0$, the computation of the jet functions becomes rather non-trivial. Restricting ourselves to the divergence in the analytic regulators, we find up to terms of $\mathcal{O}(\epsilon)$
\begin{equation}
   {\cal J}_L(b,p^\perp) = \delta\Big(b-\frac{p}{2}\Big) 
    + \frac{2C_F\alpha_s}{\pi} \left[ \frac{e^{\epsilon\gamma_E}}{\alpha-\beta}\,
    \frac{1}{b} \left( \frac{\mu}{b} \right)^{2\epsilon} 
    \bigg( 1-\frac{p^2}{4b^2} \bigg)^{-1-2\epsilon} \bigg( \frac{\nu_1^2}{b^2} \bigg)^\alpha
    \bigg( \frac{\nu_2^2}{Q^2} \bigg)^\beta + \dots \right] ,
\end{equation}
where $p=|p^\perp|$, and the dots represent non-singular terms in the analytic regulators. Performing the Fourier and Laplace transformations, the bare jet function takes the form
\begin{equation}\label{barJ1l}
\begin{aligned}
   \overline{{\cal J}}_L(\tau,z)
   &= \overline{\cal J}_L^{(0)}(\tau,z ) \\
   &\quad\times \left[ 1 + \frac{C_F\alpha_s}{\pi}\,\frac{1}{\beta-\alpha}\,
    \bigg( \frac{1}{\epsilon} + \ln\big(\mu^2\bar\tau^2 \big) 
    + 2\ln\frac{\sqrt{1+z^2}+1}{4} \bigg) \big( \nu_1^2\bar\tau^2 \big)^\alpha
    \bigg( \frac{\nu_2^2}{Q^2} \bigg)^\beta + \dots \right] ,
\end{aligned}
\end{equation}
where we have defined $\bar\tau=\tau e^{\gamma_E}$, and $\overline{\cal J}_L^{(0)}(\tau,z )$ is the tree-level jet function from (\ref{treejet}). 

\begin{figure}
\begin{center}
\begin{tabular}{ccccccc}
\includegraphics[width=0.19\textwidth]{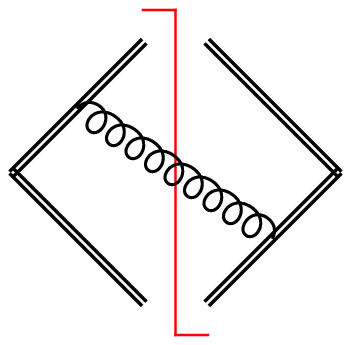} & & 
\includegraphics[width=0.19\textwidth]{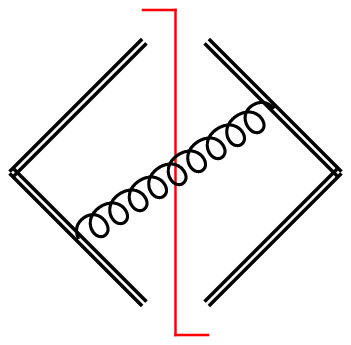} & &
\includegraphics[width=0.19\textwidth]{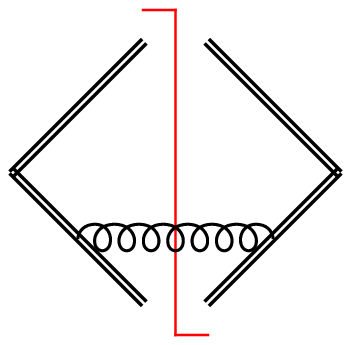} & & 
\includegraphics[width=0.19\textwidth]{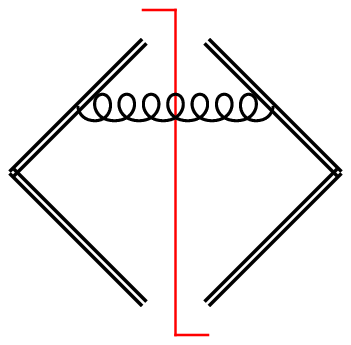}
\end{tabular}
\end{center}
\vspace{-0.5cm}
\caption{\label{fig:oneloopsoft}
Next-to-leading corrections to the soft function. The one-loop virtual diagrams are scaleless and vanish.}
\end{figure}
 
The soft function is obtained by evaluating the diagrams in Figure~\ref{fig:oneloopsoft}. The corresponding analytically-regularized Wilson-line integrals are simpler to evaluate than those for the jet function, so we are able to obtain a closed expression at one-loop order. In momentum space, we find (with $p_L=|p_L^\perp|$)
\begin{equation}
\begin{aligned}
   {\cal S}(b_L,b_R,p_L^\perp, p_R^\perp) 
   &= \delta(b_L)\,\delta(b_R)\,\delta^{d-2}(p_L^\perp)\,\delta^{d-2}(p_R^\perp) 
    + \frac{C_F\alpha_s}{\pi^{2-\epsilon}} \left( \mu^2 e^{\gamma_E} \right)^\epsilon
    \left( \frac{\nu_1^2}{Q} \right)^\alpha \left( \frac{\nu_2^2}{Q} \right)^\beta \\
   &\quad\times \frac{1}{\beta-\alpha} \left[ p_L^{-2-\alpha-\beta}\,
    \delta\Big(b_L-\frac{p_L}{2}\Big)\,\delta(b_R)\,\delta^{d-2}(p_R^\perp)
    - (L\leftrightarrow R) \right] .
\end{aligned}
\end{equation}
The Laplace and Fourier transforms can be performed explicitly, yielding
\begin{equation}
\begin{aligned}
   \overline{\cal S}(\tau_L,\tau_R,z_L,z_R) 
   &= 1 + \frac{C_F\alpha_s}{\pi}\,\frac{1}{\beta-\alpha}\,
    \frac{2^{1-\alpha-\beta-2\epsilon}\,\Gamma(-\alpha-\beta-2\epsilon)}{\Gamma(1-\epsilon)}
    \left( \mu^2 e^{\gamma_E} \right)^\epsilon \left( \frac{\nu_1^2}{Q} \right)^\alpha 
    \left( \frac{\nu_2^2}{Q} \right)^\beta \\
   &\quad\times \left[ \tau_L^{\alpha+\beta+2\epsilon}\,
    {}_2F_1\bigg(\!-\frac{\alpha+\beta+2\epsilon}{2},\frac{1-\alpha-\beta-2\epsilon}{2},
                 1-\epsilon,-z_L^2\bigg) - (L\leftrightarrow R) \right] .
\end{aligned}
\end{equation}
We now expand this expression by taking the limits $\beta\to 0$, $\alpha\to 0$, and $\epsilon\to 0$. The order in which the analytic regulators are taken to zero is arbitrary, but it is important that the limit $\epsilon\to 0$ is taken at the end, since only then the QCD result is independent of the analytic regularization. The final expression for the bare soft function is
\begin{equation}\label{barS1l}
\begin{aligned}
   \overline{\cal S}(\tau_L,\tau_R,z_L, z_R) 
   &= 1 + \frac{C_F\alpha_s}{4\pi}\,\Bigg\{ -\frac{2}{\epsilon^2}
    - \frac{2}{\epsilon}\,\ln(\mu^2 \bar{\tau}_L^2) - \ln^2(\mu^2 \bar{\tau}_L^2) \\
   &\quad\mbox{}+ 4 \left( \frac{1}{\alpha} + \ln\frac{\nu_1^2 \bar{\tau}_L}{Q} \right) 
    \left[ \frac{1}{\epsilon} + \ln(\mu^2 \bar{\tau}_L^2) + 2\ln\frac{\sqrt{1+z_L^2}+1}{4} 
    \right] \\
   &\quad\mbox{}+ 8\,\mbox{Li}_2\bigg(\! -\frac{\sqrt{1+z_L^2}-1}{\sqrt{1+z_L^2}+1} \bigg) 
    + 4\ln^2\frac{\sqrt{1+z_L^2}+1}{4} + \frac{5\pi^2}{6} - (L\leftrightarrow R) \Bigg\} \,,
\end{aligned}
\end{equation}
where $\bar\tau_{L}=  \tau_L e^{\gamma_E}$. The coefficients of the $1/\alpha$ poles are equal and opposite to those in the jet functions, see (\ref{barJ1l}), so that these divergences cancel in the product $\overline{\cal J}_L\,\overline{\cal J}_R\,\overline{\cal S}$.

Let us now use the expressions derived above to compute the differential cross section at one-loop order. In this approximation we only need the convolutions of the tree-level soft function with the one-loop jet functions and vice versa. Since the tree-level soft function involves $\delta$-functions in the transverse momenta, we only need the jet function ${\cal J}_L(b,p^\perp=0)$ given in (\ref{oneloopJ}). The resulting expression can be refactorized in the form
\begin{equation}\label{fixedsing}
   \frac{1}{\sigma_0}\,\frac{d^2\sigma}{db_L\,db_R} = H(Q^2)\,\Sigma(b_L)\,\Sigma(b_R) \,,
\end{equation}
where
\begin{equation}\label{Sigma}
   \Sigma(b) = \delta(b) + \frac{C_F\alpha_s}{4\pi}\,
   \frac{e^{\epsilon\gamma_E}}{\Gamma(1-\epsilon)}\,\frac{1}{b} 
   \left( \frac{\mu}{b} \right)^{2\epsilon} \left( 4\ln\frac{Q^2}{b^2} - 6 - 2\epsilon \right) .
\end{equation}
The dependence on the analytic regulator has indeed canceled among the jet and soft functions, and the anomalous logarithmic dependence on $Q$ is now manifest. The above expression for $\Sigma(b)$ is valid for arbitrary values of $\epsilon$. Expanding around $\epsilon=0$, and carefully treating the resulting distributions in $b$, one finds the one-loop divergence
\begin{equation}
   \Sigma(b) = \frac{C_F\alpha_s}{4\pi}\,\delta(b) \left( \frac{2}{\epsilon^2}
   - \frac{2}{\epsilon}\,\ln\frac{Q^2}{\mu^2} + \frac{3}{\epsilon} \right) 
   + \mathcal{O}(\epsilon^0)\,.
\end{equation}
This is equal to minus one half of the divergence of the bare hard function, so that the full cross section (\ref{fixedsing}) is indeed finite. In terms of renormalized objects, this result implies that the convolution of the jet and soft functions has the same anomalous dimension as the hard function, such that the cross section is renormalization-group (RG) invariant. We have used expression (\ref{fixedsing}) to compute the leading singular terms for small $b_T$ in the total broadening and reproduce the one-loop result (\ref{RB}). Having performed these one-loop checks, we now turn to the resummation of the logarithmically-enhanced corrections.

\section{Resummation}

The explicit results in the previous section show that the jet and soft functions contain divergences in the analytic regulators. These divergences cancel in the product of these functions, but they leave behind large logarithms of the momentum transfer over the broadening, which should be resummed to all orders in order to get reliable predictions. 

Interestingly, the one-loop divergences in the analytic regulators are multiplicative, such that the product 
\begin{equation}\label{Pdef}
   P(Q^2,\tau_L,\tau_R,z_L,z_R,\mu) 
   = \overline{\cal J}_L(\tau_L,z_L,\mu)\,\overline{\cal J}_R(\tau_R,z_R,\mu)\,
    \overline{\cal S}(\tau_L,\tau_R,z_L,z_R,\mu) 
\end{equation}
is finite even before the integrations over $z_L$ and $z_R$ in (\ref{factfinal}) are performed. To show that this property holds to all orders in perturbation theory, we now study a modified event shape, in which we consider the emission of two soft photons with momenta $p_{\gamma_L}$ and $p_{\gamma_R}$ in addition to the QCD partons. We then determine the thrust axis using all final-state particles, including the photons, but measure the broadening of the hadrons only. In the presence of  photons in the hemisphere, the total hadronic transverse momentum no longer adds up to zero, but will be equal and opposite to the photon momentum.

The emission of soft photons is described by QED Wilson lines, analogously to the ones appearing in the QCD soft function in (\ref{softfun}). For our purposes, it is sufficient to treat the photons classically and to consider the case where each hemisphere contains one photon with fixed momentum. The QED soft function is then given by
\begin{equation}
   S_{\rm QED}(p_{\gamma_L},p_{\gamma_R}) 
   = 4e^4\,\frac{n\cdot\bar{n}}{n\cdot p_{\gamma_L}\,\bar{n}\cdot p_{\gamma_L}}\,
    \frac{n\cdot\bar{n}}{n\cdot p_{\gamma_R}\,\bar{n}\cdot p_{\gamma_R}} \,,
\end{equation}
and it is just an overall factor multiplying the hadronic cross section. We now analyze the differential cross section
\begin{equation}\label{altfac}
\begin{aligned}
   \frac{1}{S_{\rm QED}(p_{\gamma_L},p_{\gamma_R})}\,
   \frac{d^8\sigma}{db_L\,db_R\,d^3p_{\gamma_L}\,d^3p_{\gamma_R}} 
   &\propto H(Q^2,\mu) \int\!db_L^s \int\!db_R^s \int\!d^{d-2}p^\perp_L 
    \int\!d^{d-2}p^\perp_R \\
   &\hspace{-5cm}\times {\cal J}_L(b_L-b_L^s,-p_L^\perp,\mu)\,
    {\cal J}_R(b_R-b_R^s,-p_R^\perp,\mu)\,
    {\cal S}(b_L^s,b_R^s,p_L^\perp-p_{\gamma L}^\perp,p_R^\perp-p_{\gamma R}^\perp,\mu) \,,
\end{aligned}
\end{equation}
which fulfills a factorization theorem involving exactly the same jet and soft functions as in (\ref{naivefac}), except that the soft and collinear transverse momenta now add up to minus the photon transverse momenta instead of zero. Taking the Laplace transforms in the broadenings, and Fourier transforming with respect to the photon transverse momenta $p_{\gamma L}^\perp$ and $p_{\gamma R}^\perp$, we find that the cross section is given by the same product of the Laplace and Fourier-transformed jet and soft functions as in (\ref{ff1}), but without the integrations over $x_{L,R}^\perp$. Since (\ref{altfac}) gives a physical cross section, it follows that the product
$P$ of the renormalized jet and soft functions in (\ref{Pdef}) must be well defined and independent of the analytic regulators to all orders in perturbation theory. On the left-hand side of that equation we have indicated that this product carries an anomalous dependence on the hard scale $Q$. The naive factorization formulae (\ref{naivefac}), (\ref{factfinal}), and (\ref{altfac}) thus do not achieve a proper scale separation.

The fact that the product $P$ is independent of the regulator scales $\nu_1$ and $\nu_2$ has important consequences, since the two scales enter the various functions in a very specific way. To derive these, let us send $\beta\to 0$ first. As long as $\alpha$ is non-zero the corresponding limit is regular and the jet and soft functions are $\nu_2$ independent. What remains in the individual functions is a dependence on the regulator scale $\nu\equiv \nu_1$. However, the $\nu$ independence of the product implies that $d\ln P/d\ln\nu^2=0$, where
\begin{equation}\label{lnJJS}
   \ln P = \ln\overline{\cal J}_L\big( \ln\nu^2\bar\tau_L^2;\,\tau_L,z_L,\mu \big) 
    + \ln\overline{\cal J}_R\Big( \!\ln\frac{\nu^2}{Q^2};\,\tau_R,z_R,\mu \Big) 
    + \ln\overline{\cal S}\Big( \!\ln\frac{\nu^2\bar\tau_L}{Q};\, 
    \tau_L,\tau_R,z_L,z_R,\mu \Big) \,.
\end{equation}
We indicate the explicit form of the $\nu$ dependence of the three functions in the first argument. The use of $\bar\tau_L$ rather than $\bar\tau_R$ in the first argument of the soft function is a matter of choice. In the collinear sector of left-moving particles, the logarithmic dependence on $\nu$ can for dimensional reasons only appear in the form of $\ln(\nu^2\bar\tau_L^2)$, since no other momentum scale is introduced by the $\alpha$ regulator in (\ref{analytreg}). In the collinear sector of right-moving particles, on the other hand, the induced dependence on $\nu$ is accompanied by $\bar n\cdot p_L$, because the collinear Wilson line is regularized in analogy with (\ref{wilsonreg}). By Lorentz invariance, a logarithm of the ``foreign'' momentum component $\bar n\cdot p_L$ can only appear in the form of $n\cdot p_R\,\bar n\cdot p_L=Q^2$, so that we encounter $\ln(\nu^2/Q^2)$. Finally, in the soft sector the relevant scale is always given by the geometric average of the large and small scales, yielding $\ln(\nu^2\bar\tau_L/Q)$ in the present case. 

We now call $f_L(L;\,\tau_L,z_L,\mu)=\ln\overline{\cal J}_L(L;\,\tau_L,z_L,\mu)$ and similarly for the right jet function, and denote by $f_L^{(n)}(L;\,\tau_L, z_L,\mu)$ the $n$-th derivative of this functions with respect to its first argument. The condition that the sum (\ref{lnJJS}) as well as its derivatives with respect to $\ln Q$ must be independent of the analytic regulator scale $\nu$ yields the conditions
\begin{equation}
   f_L^{(n)}\big( \ln\nu^2\bar\tau_L^2;\,\tau_L,z_L,\mu \big) 
   = \left( 2^{n-1} - 1 \right) 
    f_R^{(n)}\Big( \!\ln\frac{\nu^2}{Q^2};\,\tau_R,z_R,\mu \Big) \,; 
   \quad n\ge 2 \,.
\end{equation}
It follows that the second derivatives of the two functions must be equal to the same constant $k_2$, and all higher derivatives vanish:
\begin{equation}
\begin{aligned}
   f_L(L;\,\tau_L,z_L,\mu) 
   &= \frac{k_2(\mu)}{2}\,L^2 + k_1^L(\tau_L,z_L,\mu)\,L
    + k_0^L(\tau_L,z_L,\mu) \,, \\
   f_R(L;\,\tau_R,z_R,\mu) 
   &= \frac{k_2(\mu)}{2}\,L^2 + k_1^R(\tau_R,z_R,\mu)\,L
    + k_0^R(\tau_R,z_R,\mu) \,.
\end{aligned}
\end{equation}
We can now evaluate (\ref{lnJJS}) with the particular choice $\nu^2=Q/\bar\tau_L$, for which 
\begin{equation}
   \ln P = k_2(\mu) \ln^2(Q\bar\tau_L) 
    + \Big[ k_1^L(\tau_L,z_L,\mu) - k_1^R(\tau_R,z_R,\mu) \Big]\,\ln(Q\bar\tau_L) 
    + \dots \,,
\end{equation}
where the dots represent $Q$-independent terms. Finally, the fact that the result must be left-right symmetric implies that $k_1^L(\tau,z,\mu)+k_2(\mu)\ln(\mu\bar\tau)=-k_1^R(\tau,z,\mu)-k_2(\mu)\ln(\mu\bar\tau)\equiv-2F_B(\tau,z,\mu)$, and hence the final answer can be written in the form
\begin{equation}\label{collan}
\begin{aligned}
   \ln P 
   &= \frac{k_2(\mu)}{4}\ln^2\!\big(Q^2\,\bar\tau_L\bar\tau_R\big)
    - F_B(\tau_L,z_L,\mu)\,\ln\big(Q^2\bar\tau_L^2\big) 
    - F_B(\tau_R,z_R,\mu)\,\ln\big(Q^2\bar\tau_R^2\big) \\
   &\quad\mbox{}+ \ln W(\tau_L,\tau_R,z_L,z_R,\mu) \,,
\end{aligned}
\end{equation}
where the remainder function $W$ is independent of $Q$ and left-right symmetric.

We can gain further information by exploiting the fact that the cross section (\ref{altfac}) and hence the product $H(Q^2,\mu)\,P(Q^2,\tau_L,\tau_R,z_L,z_R,\mu)$ must be RG invariant. From the RG equation for the hard function \cite{Becher:2010tm}
\begin{equation}\label{Hevol}
   \frac{d}{d\ln\mu}\,H(Q^2,\mu) 
   = \left[ 2\Gamma_{\rm cusp}(\alpha_s)\,\ln\frac{Q^2}{\mu^2} + 4\gamma^q(\alpha_s)
    \right] H(Q^2,\mu) \,,
\end{equation}
it then follows that 
\begin{equation}
\begin{aligned}
   \frac{d}{d\ln\mu}\,k_2(\mu) &= 0 \,, \qquad\qquad
   \frac{d}{d\ln\mu}\,F_B(\tau,z,\mu) = \Gamma_{\rm cusp}(\alpha_s) \,, \\
   \frac{d}{d\ln\mu}\,W(\tau_L,\tau_R,z_L,z_R,\mu) 
   &= \Big[ 2\Gamma_{\rm cusp}(\alpha_s)\,\ln\big( \mu^2\bar\tau_L\bar\tau_R \big)
    - 4\gamma^q(\alpha_s) \Big]\,W(\tau_L,\tau_R,z_L,z_R,\mu) \,.
\end{aligned}
\end{equation}
The first equation implies that $k_2$ must be a constant. From the fact that at tree level this constant vanishes it follows that $k_2=0$ to all orders, since there would be no way to compensate the scale dependence of the coupling $\alpha_s(\mu)$. Next, using our explicit one-loop results in (\ref{barJ1l}) and (\ref{barS1l}), we find that 
\begin{equation}\label{FBres}
   F_B(\tau,z,\mu) = \frac{C_F\alpha_s}{\pi}\,\bigg[ \ln(\mu\bar\tau) 
    + \ln\frac{\sqrt{1+z^2}+1}{4} \bigg] + \mathcal{O}(\alpha_s^2) \,.
\end{equation}

We are now in the position to state the main result of this paper, which is the corrected, all-order generalization of the naive factorization theorem (\ref{factfinal}) for the Laplace-transformed double-differential cross section. Relation~(\ref{collan}), combined with the fact that $k_2=0$, implies that the anomalous dependence of the jet and soft functions on $Q$ exponentiates, and that the cross section can be refactorized in the form
\begin{equation}\label{Factfinal}
\begin{aligned}
   \frac{1}{\sigma_0}\,\frac{d^2\sigma}{d\tau_L\,d\tau_R} 
   & = H(Q^2,\mu) \int_0^\infty\!dz_L \int_0^\infty\!dz_R\,
    \big(Q^2\bar\tau_L^2\big)^{-F_B(\tau_L,z_L,\mu)}\,
    \big(Q^2\bar\tau_R^2\big)^{-F_B(\tau_R,z_R,\mu)} \\
   &\quad\times W(\tau_L,\tau_R,z_L,z_R,\mu) \,.
\end{aligned}
\end{equation}
This result contains two sources of $Q$ dependence: one arising from the hard function $H(Q^2,\mu)$, and an additional one stemming from the collinear anomaly. Once the RG equation (\ref{Hevol}) for the hard function $H(Q^2,\mu)$ is solved and the result is evaluated at a scale $\mu\sim 1/\tau_L\sim 1/\tau_R$, the above formula no longer contains any large logarithms in the perturbative series for the functions $F_B$ and $W$.

It is convenient to factor out the leading-order jet functions from the remainder function $W$ by rewriting
\begin{equation}
   W(\tau_L,\tau_R,z_L,z_R,\mu) 
   = \frac{z_L}{\left(1+z_L^2\right)^{3/2}}\,\frac{z_R}{\left(1+z_R^2\right)^{3/2}}\,
    \overline{W}(\tau_L,\tau_R,z_L,z_R,\mu) \,, 
\end{equation}
where $\overline{W}=1+\mathcal{O}(\alpha_s)$. At NLL order, we can evaluate the cross section using the tree-level result $\overline{W}=1$ and the one-loop expression for $F_B$ given in (\ref{FBres}). This yields
\begin{equation}
   \frac{1}{\sigma_0}\,\frac{d^2\sigma}{d\tau_L\,d\tau_R} 
   = H(Q^2,\mu) \left( \mu\bar\tau_L \right)^{-\eta_L} 
     \left( \mu\bar\tau_R \right)^{-\eta_R} I(\eta_L)\,I(\eta_R) \,,
\end{equation}
where
\begin{equation}
   \eta_{L,R} = \frac{C_F\alpha_s(\mu)}{\pi}\,\ln\big( Q^2\bar\tau_{L,R}^2 \big) \,,
\end{equation}
and
\begin{equation}
   I(\eta) = \int_0^\infty\!dz\,\frac{z}{\left(1+z^2\right)^{3/2}}\, 
    \bigg( \frac{\sqrt{1+z^2}+1}{4} \bigg)^{-\eta} 
%   = 4^\eta \int_0^1\!dy\,\bigg( \frac{1+y}{y} \bigg)^{-\eta}
    = \frac{4^\eta}{1+\eta}\,\,{}_2F_1(\eta,1+\eta,2+\eta,-1) \,.
\end{equation}
The function $I(\eta)$ was called $(2/\lambda)^\eta$ in \cite{Dokshitzer:1998kz}. Finally, the relevant expression for the hard function at this order reads \cite{Becher:2006mr}
\begin{equation}
\begin{aligned}
   \ln H(Q,\mu) 
   &= \frac{4C_F}{\beta_0^2}\,\bigg[
    \frac{4\pi}{\alpha_s(Q)} \left( 1 - \frac{1}{r} - \ln r \right)
    + \left( K - \frac{\beta_1}{\beta_0} \right) (1-r+\ln r) \\
   &\hspace{1.6cm}\mbox{}+ \frac{\beta_1}{2\beta_0}\,\ln^2 r 
    + \frac{3\beta_0}{2}\,\ln r \bigg] \,,
\end{aligned}
\end{equation}
where $r=\alpha_s(\mu)/\alpha_s(Q)$, $\beta_0$ and $\beta_1$ are the first two expansion coefficients of the QCD $\beta$-function, and $K=\big(\frac{67}{9}-\frac{\pi^2}{3}\big)\,C_A-\frac{20}{9}\,T_F n_f$. At NLL order we can further approximate 
\begin{equation}
   \eta_{L,R}\approx \eta 
   \equiv \frac{C_F\alpha_s(\mu) }{\pi}\,\ln\frac{Q^2}{\mu^2} \,,
\end{equation}
since $\ln(\mu\bar\tau_{L,R})$ is a small logarithm and counts as $\mathcal{O}(1)$. Then the cross section exhibits only simple power dependence on $\tau_{L,R}$, and the Mellin inversion from Laplace to $b_{L,R}$ space can be performed analytically. We finally obtain
\begin{equation}\label{nll1}
   \frac{1}{\sigma_0}\,\frac{d^2{\sigma}}{db_L\,db_R} 
   = H(Q^2,\mu)\,\frac{e^{-2\gamma_E\eta}}{\Gamma^2(\eta)}\,
    \frac{1}{b_L} \left( \frac{b_L}{\mu} \right)^{\eta} \frac{1}{b_R} 
    \left( \frac{b_R}{\mu} \right)^{\eta} I^2(\eta) \,.
\end{equation}
For the total broadening $b_T=b_L+b_R$ and the wide broadening $b_W={\rm max}(b_L,b_R)$, we find
\begin{equation}
\begin{aligned}\label{nll}
   \frac{1}{\sigma_0}\,\frac{d\sigma}{db_T} 
   &= H(Q^2,\mu)\,\frac{e^{-2\gamma_E\eta}}{\Gamma(2\eta)}\,\frac{1}{b_T} 
    \left( \frac{b_T}{\mu} \right)^{2\eta} I^2(\eta) \,, \\
   \frac{1}{\sigma_0}\,\frac{d\sigma}{db_W} 
   &= H(Q^2,\mu)\,\frac{2\eta\,e^{-2\gamma_E\eta}}{\Gamma^2(1+\eta)}\,\frac{1}{b_W}
    \left( \frac{b_W}{\mu} \right)^{2\eta} I^2(\eta) \,.
\end{aligned}
\end{equation}
In these results the resummation of the LL and NLL terms is accomplished by setting $\mu\sim b_L\sim b_R$ or $\mu\sim b_T, b_W$, respectively.

Our expressions for the cross sections in (\ref{nll}) are completely equivalent to the NLL results obtained in \cite{Dokshitzer:1998kz}, while an earlier result derived in \cite{Catani:1992jc} misses the factor $I^2(\eta)$ and is thus only correct at leading double-logarithmic order. In order to relate our expressions to the ones derived in these papers, we choose $\mu=b_T$ and rewrite the coupling constant $\alpha_s(b_T)$ at the low scale in terms of $\alpha_s(Q)$ using
\begin{equation}
   \alpha_s(b_T) = \frac{\alpha_s(Q)}{1+\omega} \left[ 1 - 
    \frac{\beta_1\,\ln(1+\omega)}{\beta_0\,(1+\omega)}\,\frac{\alpha_s(Q)}{4\pi} 
    + \dots \right] ; \qquad 
   \omega = \frac{\beta_0\alpha_s(Q)}{4\pi}\,\ln\frac{b_T^2}{Q^2} \,.
\end{equation}

\begin{figure}[t!]
\begin{center}
\begin{tabular}{cc}
\includegraphics[width=0.45\textwidth]{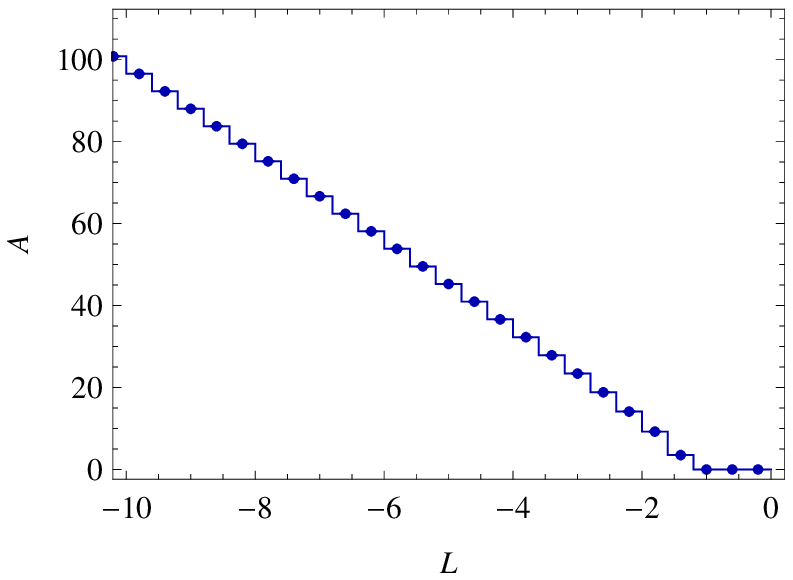} & 
\includegraphics[width=0.45\textwidth]{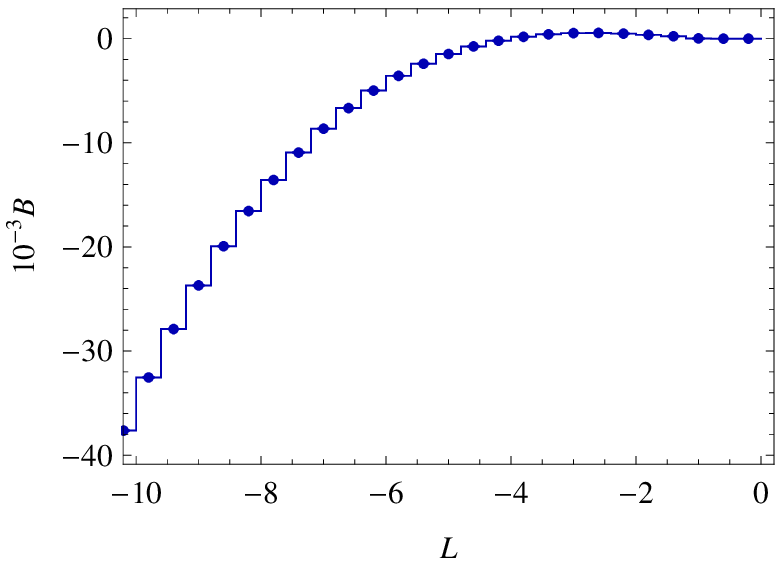} \\
\includegraphics[width=0.45\textwidth]{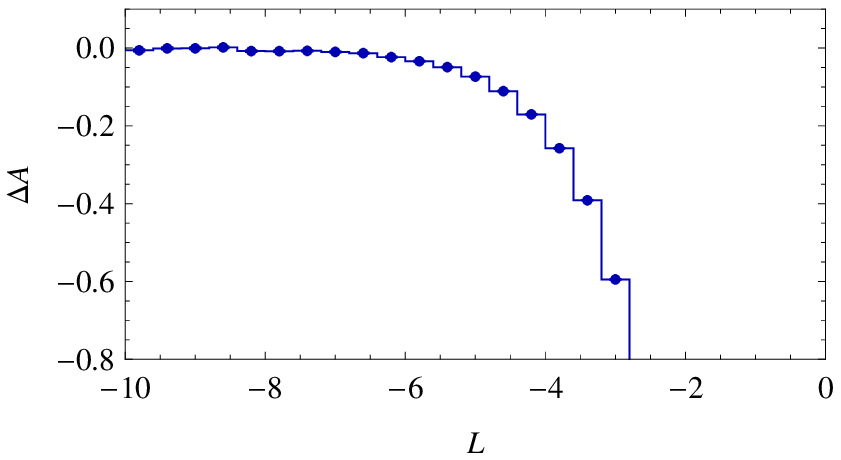} & 
\includegraphics[width=0.45\textwidth]{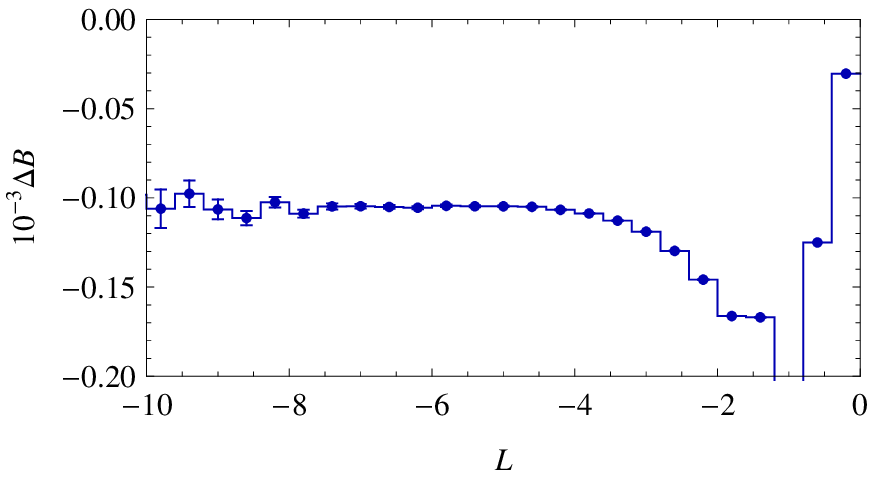}
\end{tabular}
\end{center}
\vspace{-0.5cm}
\caption{\label{event}
The upper panels show the LO and NLO coefficients in the expansion (\ref{exp2}) of the total broadening cross section as a function of $L=\ln(b_T/Q)$, as computed by the generator EVENT2 \cite{Catani:1996vz}. The lower panels show the difference of the full results to the NLL approximations.} 
\end{figure}

As a further test of our calculations, we compare our results to the fixed-order calculation of the jet broadening cross section. To our knowledge, an analytical result for this quantity is not available in the literature -- not even at $\mathcal{O}(\alpha_s)$, which is leading order (LO) for the distribution. Numerically, the cross section can be computed to next-to-leading order (NLO) using the EVENT2 generator \cite{Catani:1996vz}. The NNLO result has been obtained in \cite{GehrmannDeRidder:2007hr,Weinzierl:2009ms}. Figure~\ref{event} shows the difference between the NLO result as computed using EVENT2 and the NLL singular terms predicted by our resummation formula. In the figure we show 
\begin{equation}\label{exp2}
   \frac{b_T}{\sigma_0}\,\frac{d\sigma}{db_T} 
   = \frac{\alpha_s(Q)}{2\pi}\,A(b_T) + \bigg( \frac{\alpha_s(Q)}{2\pi} \bigg)^2 B(b_T) \,.
\end{equation}
The leading singular terms for small $b_T$, which are predicted at NLL and can be obtained by expanding (\ref{nll}), are given by
\begin{equation}
\begin{aligned}
   A^{\rm NLL}(b_T) &= C_F \left( -8L - 6 \right) , \\
   B^{\rm NLL}(b_T) &= C_F^2 \left[ 32 L^3 + 72 L^2
    + \left( 92 - \frac{40\pi^2}{3} - 64\ln^2 2 \right) L \right] \\
   &\quad\mbox{}+ C_F C_A \left[ \frac{88}{3}\,L^2
    + \left( \frac{4\pi^2}{3} - \frac{70}{9} \right) L \right] 
    + C_F T_F n_f \left( - \frac{32}{3}\,L^2 + \frac{8}{9}\,L\right) ,
\end{aligned}
\end{equation}
with $L=\ln(b_T/Q)$. At leading power in $b_T/Q$, all terms in $A(b_T)$ are predicted by our results, while the constant term in $B(b_T)$ can only be obtained by extending our analysis to NNLL order. For $b_T\to 0$, the singular terms dominate the cross section. We should thus find that the difference $\Delta A=A(b_T)-A^{\rm NLL}(b_T)$ tends to zero for $b_T\to 0$, while $\Delta B=B(b_T)-B^{\rm NLL}(b_T)$ approaches a constant in the limit of small broadening. This is indeed what is observed in Figure~\ref{event}. Let us note that a similar numerical comparison was also shown in \cite{Dokshitzer:1998kz}. However, we have run EVENT2 with a very low cutoff of $10^{-16}$ and have computed $10^8$ events using quadruple precision, which was computationally not yet feasible at the time when that paper was published.

\section{Summary and outlook}

Using effective field-theory methods, we have presented the first all-order factorization theorem for jet broadening distributions in $e^+ e^-$ collisions, in which all large logarithmic corrections arising for small broadening are resummed. Our result has several interesting features. First, the broadening receives both collinear and soft contributions, with the collinear jets in each hemisphere recoiling against the soft radiation. These recoil effects were discovered in \cite{Dokshitzer:1998kz} and become non-trivial at NLL order. Second, the individual jet and soft functions are only defined if additional regulators beyond dimensional regularization are introduced. We have used analytic regulators for that purpose. The individual functions then contain divergences in these regulators, which cancel in the product of jet and soft functions entering the formula for the differential cross section. When the regulators are removed an anomalous $Q$ dependence arises, which constitutes a new source of large logarithms. Such an anomaly also appears in transverse-momentum resummation in Drell-Yan production \cite{Becher:2010tm} and in electroweak Sudakov resummation \cite{Chiu:2007dg}. Using similar arguments as in these papers, we have shown that the associated anomalous $Q$ dependence of the broadening distribution exponentiates. 

The analytical results obtained in this paper allow us to resum logarithms to NLL accuracy, and in this way we have reproduced the results of \cite{Dokshitzer:1998kz}. In addition, we have checked the resummed expressions against the numerical fixed-order results and find that our expressions correctly reproduce the numerical results for small broadening. It would be interesting to obtain NNLL accuracy for the total and wide broadening and to combine the resummed expressions with the existing fixed-order results at NNLO obtained in \cite{GehrmannDeRidder:2007hr,Weinzierl:2009ms}. This could be used for a precise determination of $\alpha_s$ from the existing measurements of the total and wide broadening distributions. For the event-shape variable thrust, the resummation has been performed to N$^3$LL accuracy \cite{Becher:2008cf}, and this result was used for a precision determination of $\alpha_s$ in \cite{Abbate:2010xh}. The resulting value is much lower than the world average of $\alpha_s$, and it is important to cross check whether other event shapes lead to a consistent value, if only to gauge whether the uncertainty estimates are reliable. To obtain NNLL accuracy for broadening, one needs the one-loop expressions for the jet and soft functions and the two-loop expression for the anomaly function $F_B$. The one-loop soft function has already been given in this paper. The computation of the one-loop jet function is however more complicated. Finally, to obtain the two-loop anomaly coefficient one needs to compute the two-loop divergence of the soft function in the analytic regulators. We hope to report results for these quantities in a future publication.

\vspace{3mm}
\noindent{\em Note added:\/} 
While this paper was in preparation the preprint \cite{Chiu:2011qc} appeared, which also considers the resummation of jet broadening in the context of SCET. This work contains the  factorization theorem (\ref{naivefac}), but uses an alternative analytic regulator for the jet and soft functions. In contrast to our work, the regularization is introduced directly in SCET instead of QCD, and it is not clear to us whether QCD is recovered once this regulator is removed. More importantly, the result for the total broadening presented in \cite{Chiu:2011qc} does not account for the recoil effects discussed in \cite{Dokshitzer:1998kz} and is therefore missing the factor $I^2(\eta)$ in (\ref{nll1}) and (\ref{nll}). It is thus incorrect beyond the  leading double-logarithmic order.

\vspace{3mm}
\noindent{\em Acknowledgments:\/} 
We are grateful to Gavin Salam and Giulia Zanderighi for useful discussions and help with EVENT2. T.B.\ and G.B.\ are supported in part by SNSF and ``Innovations- und Kooperationsprojekt C-13'' of SUK. The research of M.N.\ is supported in part by BMBF grant 05H09UME, DFG grant NE 398/3-1, and the Research Center ``Elementary Forces and Mathematical Foundations''.

\end{document}